\documentstyle[12pt]{article}\topmargin=-25mm\leftmargin=25mm
\begin{document}
\bibliographystyle{unsrt}\begin{center}{\large\bf
An Approach to Maxwell Equations in Unifromly Accelerated Spherical
Coordinates by Newman-Penrose Method}\\[5mm]
Z.\,Ya.\,Turakulov\\{\it Institute of Nuclear Physics\\
Ulugbek, Tashkent 702132, Rep. of Uzbekistan, CIS\\
(e-mail:zafar@.suninp.tashkent.su)}\end{center}

Variables are separated in Maxwell equations by the Newman-Penrose method
of isotropic complex tetrade in the uniformly accelerated spherical
coordinate system. Particular solutions are obtained in terms of spin 1
spherical harmonics. PACS: 03.50.De\newpage\section{Introduction}

The uniformly accelerated spherical coordinates have been introduced in our
work \cite{doesn't}. It turns out that the Newman-Penrose method of isotropic
complex tetrade \cite{n-p} works in this coordinate system. The aim of the
present work is to demonstrate applying the method in the coordinate system and
obtain the general solution to Maxwell equations in these coordinates.

Although a comprehnsive formulation of the method have been published
\cite{g-mf-n-r-s}, in order to apply it in a new coordinate system we have to
start with composing the tetrade and thus to repeat the whole exposition. Thus,
the method is to be actually recovered in details.\section{The isotropic complex
tetrade}

The metric for the uniformly accelerated spherical coordinates
$\{\xi,u,v,\varphi\}$ has the form \cite{doesn't}
\begin{equation}ds^2=a^2{{\sinh^2ud\xi^2-du^2-dv^2-\sin^2v d\varphi^2}
\over{(\cosh u+\cos v)^2}}.\end{equation}
To compose the isotropic complex tetrade we first solve the Hamolton-Jacobi
equation for isotropic lines in this metric:
$$(\cosh u+\cos v)^2\left[\frac{1}{\sinh^2u}\left(
\frac{\partial S}{\partial\xi}\right)^2-
\left(\frac{\partial S}{\partial u}\right)^2-
\left(\frac{\partial S}{\partial v}\right)^2+\frac{1}{\sin^2v}
\left(\frac{\partial S}{\partial\varphi}\right)^2\right]=0.$$
Substituting $S=E\xi+U(u)+V(v)+M\varphi$ yields a separated equation:
$$E^2\sinh^{-2}u-U'^2=V'^2-M^2\sin^{-2}=L^2.$$

The two congruences of isotropic rays to be used below for composing the tetrade
corresponds to the values $E=1$, $L=M=0$ and $M=1,\enskip L=E=0$ of the
constants. They give respectively $$U'=\pm\sinh^{-1}u,\enskip V'=0$$and
$$U'=0,\enskip V'=\pm\imath\sin^{-1}v.$$
The isotropic co-vectors with these two congruences are normalized with respect
to the metric $$ds^2=a^2(\sinh^2ud\xi^2-du^2-dv^2-\sin^2v d\varphi^2)$$
obtained from the metric (1) by an apparent conformal transformation. It is
convenient to do so because Maxwell equations are forminvariant under conformal
transormations. After the normalization procedure the tetrade appears in the
form:\begin{equation}
\kappa=\sinh ud\xi-du,\enskip\lambda=\sinh ud\xi+du\end{equation}
$$\mu=dv+\imath\sin vd\varphi,\enskip\nu=dv-\imath\sin vd\varphi$$
The co-vectors constitute a normalized orthogonal frame with only two non-zero
scalar products:$<\kappa,\lambda>=<\mu,\nu>=1$. The reciprocal relations are:
\begin{equation}d\xi=\frac{\kappa+\lambda}{2\sinh u},
\enskip du=\frac{\lambda-\kappa}{2}\end{equation}$$
dv=\frac{\mu+\nu}{2}\enskip d\varphi=\imath\frac{\mu-\nu}{2\sin v}$$
The vector frame dual to the co-vector frame (1) is the following:
\begin{equation}
\vec k=\frac{1}{2}\left(\frac{1}{\sinh u}\frac{\partial}{\partial\xi}-
\frac{\partial}{\partial u}\right),\enskip
\vec l=\frac{1}{2}\left(\frac{1}{\sinh u}\frac{\partial}{\partial\xi}+
\frac{\partial}{\partial u}\right),\end{equation}$$
\vec m=\frac{1}{2}\left(\frac{\partial}{\partial v}-
\frac{\imath}{\sin v}\frac{\partial}{\partial\varphi}\right)\enskip
\vec n=\frac{1}{2}\left(\frac{\partial}{\partial v}+
\frac{\imath}{\sin v}\frac{\partial}{\partial\varphi}\right)$$

Consider the following 2-forms $\Phi^1,\Phi^2,\Phi^3$:
\begin{equation}\Phi^1=\kappa\wedge\mu,\enskip\Phi^2=\frac{1}{2}
(\kappa\wedge\lambda+\mu\wedge\nu),\enskip\Phi^3=\lambda\wedge\nu
\end{equation}Evaluating them due to the relations (2) gives:
\begin{equation}\Phi^1=\sinh ud\xi\wedge dv+
\imath\sinh u\sin vd\xi\wedge d\varphi-du\wedge dv+
\imath\sin vd\varphi\wedge du\end{equation}$$\Phi^2=
\sinh ud\xi\wedge du-\imath\sin v dv\wedge d\varphi$$ $$
\Phi^3=\sinh ud\xi\wedge dv-
\imath\sinh u\sin vd\xi\wedge d\varphi+du\wedge dv+
\imath\sin vd\varphi\wedge du$$
It is seen that the frame of 2-forms $\{\Phi^a\}$ is self-dual:
\begin{equation}{}^*\Phi^a=\imath\Phi^a.\end{equation}
As these three 2-forms are linearly independent they
constitute a complete frame.

Exterior derivatives of $\Phi^a$'s evaluated from the equations (5) with
inserting the expressions (3) and rewritten in terms of the tetrade (1) and
the frame $\{\Phi^a\}$' are:
\begin{equation}d\Phi^1=\frac{1}{2}\coth u\lambda \wedge\Phi^1+
\cot v\nu\wedge\Phi^1,\enskip d\Phi^2=0\end{equation}
$$d\Phi^3=-\frac{1}{2}\coth u\kappa\wedge\Phi^3+
\cot v\mu\wedge\Phi^3.$$\section{Reduction of Maxwell equations}

An arbitrary 2-form of strengths of electromagnetic field $E$ can be represented
as a expansion in the frame of $\Phi^a$'s:
\begin{equation}E=F\Phi^1+G\Phi^2+H\Phi^3\end{equation}
with $F$, $G$ and $H$ being arbitrary scalar functions. Since, due to the
equations (7) the 2-form $E$ is self-dual Maxwell equations are reduced to one
equation $dE=0$:\begin{equation}0=dE=
(\vec l\circ F)\lambda\wedge\Phi^1+(\vec n\circ F)\nu\wedge\Phi^1+
(\vec k\circ H)\kappa\wedge\Phi^3+(\vec m\circ H)\mu\wedge\Phi^3+
\end{equation}$$
(\vec k\circ G)\kappa\wedge\Phi^2+(\vec l\circ G)\lambda\wedge\Phi^2
(\vec m\circ G)\mu\wedge\Phi^2+(\vec n\circ G)\nu\wedge\Phi^2+
Fd\Phi^1+Hd\Phi^3$$
where action of vectors on scalars is the same as that of differential operators
(4). It is convenient to employ ambiguity of expression of 3-forms as exterior
products of the tetrade elements and $\Phi^a$'s to eliminate the 2-form
$\Phi^2$: $$\kappa\wedge\Phi^2=\frac{1}{2}\nu\wedge\Phi^1,\enskip
\lambda\wedge\Phi^2=-\frac{1}{2}\mu\wedge\Phi^3,$$
$$\mu\wedge\Phi^2=-\frac{1}{2}\lambda\wedge\Phi^1,\enskip
\nu\wedge\Phi^2=\frac{1}{2}\kappa\wedge\Phi^3.$$

After eliminating 3-forms containing $\Phi^2$, colleting similar terms,
inserting the expressions (8) and annulating the common factors one can rewrite
the equation (10) in the form
$$\left(\frac{\partial}{\partial u}+
\frac{1}{\sinh u}\frac{\partial}{\partial\xi}+\coth u\right)F=
\frac{1}{2}\left(\frac{\partial}{\partial v}-
\frac{\imath}{\sin v}\frac{\partial}{\partial\varphi}\right)G$$
$$\left(\frac{\partial}{\partial v}+
\frac{\imath}{\sin v}\frac{\partial}{\partial\varphi}+\cot v\right)F=
-\frac{1}{2}\left(\frac{\partial}{\partial u}-
\frac{1}{\sinh u}\frac{\partial}{\partial\xi}\right)G$$
$$\left(\frac{\partial}{\partial u}-
\frac{1}{\sinh u}\frac{\partial}{\partial\xi}+\coth u\right)H=
-\frac{1}{2}\left(\frac{\partial}{\partial v}+
\frac{\imath}{\sin v}\frac{\partial}{\partial\varphi}\right)G$$
$$\left(\frac{\partial}{\partial v}-
\frac{\imath}{\sin v}\frac{\partial}{\partial\varphi}+\cot v\right)H=
\frac{1}{2}\left(\frac{\partial}{\partial u}+
\frac{1}{\sinh u}\frac{\partial}{\partial\xi}\right)G$$
To accomplish the further reduction we put:\begin{equation}
F=f_+e^{\imath(k\xi+m\varphi)}+f_-e^{-\imath(k\xi+m\varphi)}\end{equation}
$$\frac{1}{2}G=ge^{\imath(k\xi+m\varphi)}+ge^{-\imath(k\xi+m\varphi)}$$
$$H=f_-e^{\imath(k\xi+m\varphi)}+f_+e^{-\imath(k\xi+m\varphi)}$$
and have two coinciding pairs of equations which are\begin{equation}
\left(\frac{\partial}{\partial u}+\coth u\pm
\frac{\imath k}{\sinh u}\right)f_\pm=
\left(\frac{\partial}{\partial v}\pm\frac{m}{\sin v}\right)g\end{equation}
$$\left(\frac{\partial}{\partial v}+\cot v\pm
\frac{m}{\sin v}\right)f_\pm=
\left(\frac{\partial}{\partial u}\mp\frac{\imath k}{\sinh u}\right)g$$
This system can be solved first for the functions $f$ and
then the function $g$ can be found from these equations.\section{Variables
separation and explicit form of the scalar functions}

The equations (11) reduce to the following equation for the functions
$f_\pm$:$$\left[\left(\frac{\partial^2}{\partial u^2}+
\coth u\frac{\partial}{\partial u}+
\frac{k^2\mp\imath k\cosh u+1}{\sinh^2u}\right)+
\left(\frac{\partial^2}{\partial v^2}+
\cot v\frac{\partial}{\partial v}+
\frac{m^2\pm m\cos v-1}{\sin^2v}\right)\right]f_\pm=0.$$
Taking the function to be found in factorized form
\begin{equation}f_\pm=U_\pm(u)V_\pm(v)\end{equation}
Substituting this separates the equation and gives:
$$U''_\pm+U'_\pm\coth u+\frac{k^2\pm\imath k\cosh u+1}{\sinh^2 u}
=l(l+1)U_\pm$$
$$V''_\pm+V'_\pm\cot v-\frac{m^2\pm m\cos v+1}{\sin^2 v}=-l(l+1)V_\pm.$$
Solutions of these equations are known as spin 1 spherical
harmonics \cite{g-mf-n-r-s}:$$U_\pm(u)={}_1S_{l\enskip\imath k}(u)
\enskip V_\pm(v)={}_1S_{lm}(v).$$Due to the equations (11) the function $g$
is$$g(u,v)=\frac{1}{l^2(l+1)^2}\left(\frac{\partial}{\partial u}+
\coth u\pm\frac{\imath k}{\sinh u}\right){}_1S_{l\enskip\imath k}(u)
\left(\frac{\partial}{\partial v}+
\cot v\pm\frac{m}{\sin v}\right){}_1S_{lm}(v).$$
Substituting this into the equations (13) and, further, into equations (11)
together with the equations (9) and (6) one obtains patricular solutions of
Maxwell equations, forming a complete orthogonal basis in the functional space.
In the case $m=0$ one obtains the expansion found in our work \cite{doesn't}.
\end{document}